\documentclass[a4paper,11pt]{article} 
\usepackage[utf8]{inputenc} 
\usepackage[T1]{fontenc} 
\usepackage[english]{babel} 
\usepackage{amsmath, amssymb} 
\usepackage{graphicx} 
\usepackage{float} 
\usepackage{booktabs}
\usepackage{enumitem}
\usepackage{listings} 
\usepackage{xcolor} 
\usepackage{placeins}
\usepackage{siunitx}
\usepackage{algorithm} 
\usepackage{algpseudocode}
\usepackage[ backend=biber, style=numeric, sorting=nyt ]{biblatex} 

\title{An Information-Geometric Framework for Bayesian Credit Risk Monitoring}

\author{ Lorenzo Quirini\thanks{Head of \emph{Innovazione del Credito} at Banca Monte dei Paschi di Siena. The views expressed in this paper are solely those of the author and do not represent the views, positions, or policies of the author's employing bank. Any errors are the author's own.}}

\date{August 2026}

\begin{document}

\maketitle

\begin{abstract}
We propose an information-geometric framework for credit risk monitoring in which a bank’s knowledge of a borrower is represented by a posterior distribution over latent dimensions of creditworthiness and financial fragility. Under a linear-Gaussian specification, Bayesian updating maps observed behavioural scores into Gaussian posterior beliefs, which form a statistical manifold endowed with the Fisher information metric. In the common-covariance case, the induced geometry reduces to the Mahalanobis metric on posterior means, while 
\mbox{borrower-specific} covariance matrices allow informational distances to account for both expected risk and assessment uncertainty. Simulation exercises illustrate how Kullback-Leibler and Jeffreys divergences can be used to compare portfolio segments. The framework provides a geometric interpretation of credit monitoring as the evolution of posterior beliefs over borrower risk.
\end{abstract}

\setcounter{tocdepth}{1}
\tableofcontents 
\section{Introduction}

Credit risk is interpreted as a subjective probability distribution characterizing the financial intermediary. This distribution evolves on a statistical manifold as new information becomes available. This approach differs from the prevailing view, which treats credit risk as a pointwise probability associated with the borrower.

In the proposed perspective, credit risk is therefore not an objective property of the borrower, but rather a subjective distribution held by the financial intermediary over the borrower's latent state. The starting assumption is that credit risk, whether referring to a household or a firm, can be reduced to two fundamental latent dimensions: \textbf{creditworthiness} and \textbf{financial fragility}.

\textbf{Creditworthiness} and \textbf{Fragility} are continuous latent factors associated, respectively, with the willingness and the ability of the borrower to fulfil the financial obligations undertaken.

This distinction is part of standard knowledge in banking practice. One may refer, for instance, to the Anglo-Saxon paradigm of the 3\textit{C}s (\textit{Character, Capacity, Collateral}) as a guideline for assessing credit-granting decisions.

For an introduction to credit scoring, we refer some classical textbooks like \cite{Thomas2002} or \cite{Anderson2007}.

These latent variables, denoted in what follows by the symbol \(\theta\), do not represent directly observable quantities, nor uniquely measurable economic parameters. They are abstract concepts, widely used in banking practice to summarize the borrower's ability and willingness to meet financial obligations.

They are economic quantities that the bank would like to know, but which are neither directly observable nor fully knowable. They should therefore be regarded as inferential constructs, whose meaning is defined by the subjective distribution held by the financial intermediary over the creditworthiness and financial fragility of its customers.

The focus is not on estimating a ``true'' value of the latent variables, but rather on studying the evolution of the posterior distributions that describe the bank's state of knowledge.

What the lending institution normally observes are credit or financial fragility \textit{scores}. By using these \textit{scores} as summary information, the bank assumes that they act as sufficient statistics for customer profiling. A credit \textit{score} is generally related to the probability of a specific event occurring, such as an overdraft lasting more than 90 days. The estimation of this probability is conditional on the knowledge of a wide range of variables, such as delays in financial obligations at the system level, frequency and amount of overdrafts, number of recent credit inquiries, use of revolving credit facilities, and so on.

In the case of new credit originations, in addition to behavioural data obtained from the Central Credit Register of the Bank of Italy, from private credit information providers, or from public data sources, applicant-specific socio-economic variables are often used, such as job seniority, income, occupation, type of employment, and so forth. In the case of firms, additional information can be inferred from financial statements, whereas for sole proprietorships it is often derived from the limited income information reported in tax returns.

The financial fragility \textit{score} may also be constructed, possibly using unsupervised techniques, from variables such as instalment-to-income ratio, debt-to-income ratio, debt-to-\textit{plafond} ratio, and similar indicators.

The contribution of this note is threefold. First, we formulate credit monitoring as Bayesian updating of a posterior distribution over latent creditworthiness and fragility. Second, we show that, under a linear-Gaussian specification, posterior beliefs define points on a Gaussian statistical manifold endowed with the Fisher information metric. Third, we derive segment-level informational distances based on Kullback-Leibler and Jeffreys divergences and illustrate their interpretation through simulation exercises.

The paper proceeds by introducing a Bayesian model aimed at estimating the posterior distributions of the latent variables \textbf{Creditworthiness} and \textbf{Fragility}, conditional on the initial information state \(z\) associated with each borrower (Section 2). Subsequently, in Section 3, these probability distributions are interpreted as objects belonging to a Riemannian manifold, whose metric is induced by the Fisher information matrix. In a first step, the case in which the bank's uncertainty about the latent variables is constant across the population is considered.

Section 4 illustrates the results of the previous sections through a simulation exercise. 
Section 5 generalizes the geometry to the case in which the bank's uncertainty varies from borrower to borrower. 
The following section, Section 6, revisits the numerical example by adapting it to the setting in which the bank's prior informational state varies across borrowers both in mean and in covariance.
Discussions and empirical outlook are drawn in Section 7.
Section 8 concludes.

\subsection{Related Literature}

This paper lies at the intersection of credit risk modelling, Bayesian inference, and information geometry. Its applied motivation comes from the credit process of a financial institution, in which borrowers are first selected at origination and subsequently managed through monitoring activities. Classical references on credit scoring, credit risk assessment, and portfolio monitoring include \cite{Anderson2007}, \cite{Thomas2002}, and \cite{Thomas2009}.

The probabilistic structure of the model is Bayesian. Borrowers are described through latent variables, and the information available to the bank is represented by posterior distributions updated after the observation of monitoring scores. For the general Bayesian framework, we refer to \cite{Gelman2013}. 

The geometric component of the paper is based on information geometry. The starting point is the seminal work of Rao \cite{Rao1945}, in which the Fisher information matrix is interpreted as a Riemannian metric on a statistical model. A modern introduction and reinterpretation of this perspective can be found in \cite{Nielsen2020}. The general theory of statistical manifolds, natural and expectation coordinates, dual connections, and divergence functions is developed in \cite{Amari2016}. The broader mathematical background on Riemannian geometry, including metrics, geodesics, and curvature, can be found in \cite{DoCarmo1992}.

\section{The Bayesian model}

We adopt a Bayesian approach to derive the distributions of interest, following the classical framework described in \cite{Gelman2013}.

Each borrower is characterized by a set of variables that profile the borrower at the initial stage of the credit relationship.

We denote by \(z\) the set of such characteristics, for example those collected by the bank for the computation of an origination \textit{score}. The prior distribution is a density \(p(\theta \mid z)\), which represents the lender's knowledge of the borrower derived from the initially available characteristics. In this way, each individual borrower is associated with a specific \textit{prior}. In other words, the \textit{prior} represents the initial way in which the financial institution views its borrower.

Once the borrower has been acquired, the bank starts the subsequent monitoring process in order to evaluate possible commercial development actions or corrective interventions.

This activity determines the mechanism that generates behavioural monitoring \textit{scores}, namely the variables \(x\) that are actually observable. To fix ideas, we assume that \(x\) consists of a pair of \textit{scores}, for instance a behavioural credit score \(c\) and a financial fragility score \(f\). This process is governed by a likelihood function \(p(x \mid \theta,z)\). The likelihood describes the mechanism through which the observed \textit{scores} are produced.

The posterior distribution \(p(\theta \mid x,z)\) is thus defined, and represents the updated knowledge of the borrower held by the financial institution.

Knowledge of the posterior distributions allows one to use the standard tools of information geometry, in line with the framework of Amari \cite{Amari2016}. 
Under the assumption that these distributions belong to the class of exponential-family densities, the corresponding potential function \(\psi\), also known as the log-partition function, is well defined. The whole geometry, including dual coordinates, the Fisher metric, and divergences, derives from \(\psi\).

Having established this framework, we construct a first model in order to show how the geometry of posterior distributions can be defined from the knowledge of the \textit{priors}. In particular, we show that, by suitably choosing the \textit{priors} and the likelihood function, the posterior distribution belongs to the class of exponential-family densities.

To this end, we assume that the \textit{prior} is a normal distribution whose mean vector \(m\) and variance-covariance matrix \(P\) depend on \(z\).

\begin{equation}
\theta \mid z \sim N\bigl(m(z),P(z)\bigr).
\end{equation}

The function \(m(z)\) represents the initial mean knowledge.

The variance-covariance matrix \(P(z)\) may take the following form:
\begin{equation}
P(z)=
\begin{pmatrix}
e^{u(z)} & \rho(z)e^{\frac{u(z)+v(z)}{2}} \\[0.3cm]
\rho(z)e^{\frac{u(z)+v(z)}{2}} & e^{v(z)}
\end{pmatrix},
\qquad |\rho(z)| < 1.
\end{equation}

where:
\begin{itemize}
    \item \(u(z)\) governs the uncertainty about Creditworthiness;
    \item \(v(z)\) governs the uncertainty about Fragility;
    \item \(\rho(z)\) expresses the correlation between these latent variables.
\end{itemize}

In this way, in addition to differing in the mean position of their prior distributions, borrowers may also differ in the degree of uncertainty assigned by the intermediary to the latent variables \textbf{Creditworthiness} and \textbf{Fragility}, as well as in the dependence structure between them.

In other words, two borrowers may have the same \(m(z)\), but different variance-covariance matrices \(P(z)\). This reflects the different degree of confidence that the bank associates with the relationship with each borrower.

It should be noted that each individual borrower is identified by five parameters: the two components of \(m(z)\) and the three independent components of \(P(z)\).

As far as the likelihood is concerned, we assume that, conditionally on \(\theta \mid z\), the variable \(x=(c,f)\) is normally distributed. In this setting, the \textit{scores}, namely the observable variables, are realizations of \(p(x \mid \theta,z)\), that is,
\[
x \mid \theta,z \sim N(A\theta,\Sigma),
\]
where \(\theta(z)\) denotes the latent state associated with the prior information \(z\), \(A\) is a \(2\times 2\) matrix, and \(\Sigma\) is the variance-covariance matrix.

The system belongs to the class of linear-Gaussian models. Under this specification, the prior and likelihood are conjugate, and the posterior distribution of the latent state remains Gaussian.

The matrix \(A\) describes how the latent states affect the observed scores. In particular, \(a_{11}\) measures the effect of creditworthiness on the behavioural credit score, \(a_{12}\) measures the effect of financial fragility on the behavioural credit score, \(a_{21}\) measures the effect of creditworthiness on the financial fragility score, and \(a_{22}\) measures the effect of financial fragility on its corresponding score.

Under these specifications, Bayesian updating implies that the posterior density of \(\theta\) is also Gaussian, with mean \(m_1\) and variance-covariance matrix \(P_1\).

For the posterior density, the following expressions are obtained for \(P_1(z)\) and \(m_1(z,x)\):
\begin{equation}
P_1(z)=
\left(
P^{-1}(z) + A^{\top}\Sigma^{-1} A
\right)^{-1}.
\end{equation}

\begin{equation}
m_1(z,x)=
P_1(z)
\left(
P^{-1}(z)m(z)
+
A^\top\Sigma^{-1}x
\right).
\end{equation}

The posterior distributions therefore represent the state of knowledge that the bank associates with each borrower after the monitoring process.

In general, monitoring activity is carried out periodically, for example on a monthly basis. In this case, one obtains families of posterior distributions in which the \textit{priors} are the posterior distributions from the previous period, while the likelihood function is determined by the behavioural monitoring \textit{scores} \(x\) observed at each subsequent date.

\section{The Geometry Associated with Posterior Distributions}

The framework described so far changes the perspective from which credit risk is interpreted. Traditionally, credit risk is described in terms of \textit{scores}, \textit{PD}, or \textit{ratings}. In this model, the fundamental object becomes the degree of \textbf{knowledge} held by the lending institution. The knowledge of the financial institution evolves on a statistical manifold whose points are posterior densities. The observed \textit{scores} are the events that induce a displacement on the manifold.

Our manifold is therefore
\[
M=\left\{N\bigl(m_1(z,x),P_1(z)\bigr)\right\},
\]
where \(m_1(z,x)\) and \(P_1(z)\) are, respectively, the posterior mean and the posterior variance-covariance matrix. In this model, the posterior distribution associated with the \(i\)-th borrower is represented by a ``point''
\[
N\bigl(m_i(z,x),P_i(z)\bigr).
\]

Equivalent descriptions of the same Gaussian distribution can be given using different coordinate systems, for instance expectation coordinates, used so far, and natural coordinates.

The natural coordinates are related to the expectation coordinates by:
\begin{equation}
\Lambda(z)=P_1^{-1}(z)
\end{equation}
and
\begin{equation}
\eta(z)=P_1^{-1}(z)m_1(z,x).
\end{equation}

In the literature, \(\Lambda(z)\) is also referred to as the precision matrix.

Using natural coordinates, the relationship between \(\Lambda(z)^{\mathrm{prior}}\) and \(\Lambda(z)^{\mathrm{posterior}}\) becomes:
\begin{equation}
\Lambda(z)^{\mathrm{posterior}}
=
\Lambda(z)^{\mathrm{prior}}
+
A^{\top}\Sigma^{-1}A.
\end{equation}

Analogously, the relationship between \(\eta(z)^{\mathrm{prior}}\) and \(\eta(z)^{\mathrm{posterior}}\) becomes:
\begin{equation}
\eta^{\mathrm{posterior}}(z,x)
=
\eta^{\mathrm{prior}}(z)
+
A^{\top}\Sigma^{-1}x.
\end{equation}

Rewriting the relationship between \textit{prior} and \textit{posterior} in natural coordinates shows that, in this coordinate system for posterior distributions, information is additive.

It should be kept in mind that the \textit{prior} represents what the financial institution knows before observing the \textit{scores} obtained through the monitoring process. The natural coordinate \(\Lambda(z)^{\mathrm{prior}}\) indicates how certain the lending institution is about such knowledge, while \(\eta(z)^{\mathrm{prior}}\) represents the direction of this belief, taking into account both the mean position and the precision.

When the \textit{scores} are computed, they add information: the precision increases, or is at least updated, and the position of the belief changes.

Since the posterior distributions are Gaussian, they belong to the class of exponential-family densities. It is therefore possible to derive the geometric quantities of interest, such as the metric and the divergence, by computing the potential function \(\psi\).

The expression of the potential function \(\psi\) for a \(d\)-dimensional Gaussian distribution expressed in natural coordinates is well known. To avoid overburdening the notation, \(\psi\) depends on \(z\) and \(x\) through \(\eta\) and \(\Lambda\).

\begin{equation}
\psi(\eta,\Lambda)
=
\frac12\,\eta^{T}\Lambda^{-1}\eta
-\frac12\,\log\!\bigl|\Lambda\bigr|
+\frac d2 \log(2\pi).
\end{equation}

The term \(\frac d2 \log(2\pi)\) is constant. In our case, since we work with creditworthiness and fragility, we have \(d=2\).

This function allows one to recover all the objects of information geometry. First, the gradient with respect to the natural coordinates gives the expectation parameters, that is, the dual coordinates. Second, the Hessian of \(\psi\) is the Fisher metric. Finally, the Bregman divergence associated with \(\psi\) coincides with the Kullback-Leibler divergence between two distributions belonging to the same exponential family.

With reference to the Kullback-Leibler divergence, recall that for two distributions \(p_i\) and \(p_j\) defined on a common support \(X\), it is given by:
\begin{equation}
D_{\mathrm{KL}}\!\left(p_i \,\|\, p_j\right)
=
\int p_i(x)\,
\log\!\left(
\frac{p_i(x)}{p_j(x)}
\right)\,dx.
\end{equation}

In the Gaussian case, the potential function \(\psi\) depends both on the vector \(\eta\) and on the matrix \(\Lambda\). Therefore, the Hessian matrix has a block structure, involving derivatives with respect to \(\eta\), derivatives with respect to \(\Lambda\), and mixed derivatives.

In order to derive the geometric properties of the manifold in a simple way, we initially consider the submanifold of Gaussian distributions with fixed covariance, namely the case in which \(P_1\) is constant. This assumption holds when one works in the subspace parametrized by \(\eta\), assuming that \(\Lambda\) is constant for every borrower.

We therefore consider the submanifold of Gaussian distributions obtained by keeping the precision matrix \(\Lambda\) fixed. This submanifold is parametrized exclusively by the natural coordinates \(\eta\) and has dimension two.

From a credit-risk perspective, this assumption means that the bank differentiates borrowers only in terms of their mean values and assigns the same level of uncertainty to every borrower.

This assumption will be removed in Section 5, where these results are generalized.

We now compute the gradient of \(\psi\) with respect to \(\eta\):
\begin{equation}
\frac{\partial \psi}{\partial \eta}
=
\Lambda^{-1}\eta.
\end{equation}

Since \(\Lambda^{-1}=P_1\) and \(\eta=\Lambda m_1\), it follows that:
\begin{equation}
\frac{\partial \psi}{\partial \eta}
=
m_1.
\end{equation}

As expected, the dual coordinates are precisely the mean parameters.

To obtain the Fisher metric \(g\), one has to compute the Hessian of \(\psi\). Since \(P_1\) is constant, \(\Lambda\) is the same for every borrower. Differentiating once again with respect to \(\eta\), we obtain:
\begin{equation}
\frac{\partial^2 \psi}{\partial \eta \, \partial \eta^{\top}}
=
\Lambda^{-1}.
\end{equation}

The Fisher metric in the subspace of natural coordinates \(\eta\) is therefore:
\begin{equation}
g_\eta=P_1.
\end{equation}

In this case, the Fisher metric coincides with the posterior variance-covariance matrix. It follows that the structure of uncertainty directly defines the geometry of the space of posterior probability distributions. In other words, uncertainty becomes geometry.

This means that, when the covariance matrix is fixed, the manifold is flat in the system of natural coordinates. The metric \(g=P_1\) does not depend on the borrower's \textit{score}; it depends only on the uncertainty that the lending institution assigns to the latent variables \textbf{Creditworthiness} and \textbf{Fragility}.

In other words:
\begin{itemize}
    \item if the bank is highly confident, corresponding to a small covariance matrix, distances on the manifold are more ``rigid'';
    \item if the bank is highly uncertain, corresponding to a large covariance matrix, the same variations in \textbf{Creditworthiness} and \textbf{Fragility} correspond to smaller geometric distances.
\end{itemize}

Since the Fisher metric on the submanifold characterized by constant \(\Lambda\) is
\begin{equation}
g_\eta=\Lambda^{-1}=P_1,
\end{equation}
the line element is
\begin{equation}
ds^2=d\eta^\top g_\eta d\eta=d\eta^\top P_1 d\eta.
\end{equation}

Since
\[
\eta=\Lambda m_1,
\qquad
d\eta=\Lambda\,dm_1,
\]
we obtain
\[
ds^2=(\Lambda dm_1)^\top P_1(\Lambda dm_1).
\]

Because \(P_1=\Lambda^{-1}\) and \(\Lambda\) is symmetric, it follows that
\[
ds^2=dm_1^\top \Lambda dm_1.
\]

The line element therefore coincides with the Mahalanobis metric associated with the precision matrix \(\Lambda\).

The geometric distance induced by the metric therefore coincides with the Mahalanobis distance:
\begin{equation}
d_g(p_i,p_j)
=
d_M(p_i,p_j)
=
\sqrt{
(m_{1,j}-m_{1,i})^\top
\Lambda
(m_{1,j}-m_{1,i})
}.
\end{equation}

The fixed-covariance submanifold is an affine, or e-flat, manifold in the natural coordinates \(\eta\). Thus, all borrowers live on the same flat manifold and differ only in their position.

For two Gaussian distributions with common covariance 
$P_1$, the Kullback-Leibler divergence is:

\begin{equation}
D_{\mathrm{KL}}(p_i\|p_j)
=
\frac12
(m_{1,j}-m_{1,i})^\top
\Lambda
(m_{1,j}-m_{1,i})
=
\frac12 d_M^2(p_i,p_j).
\end{equation}

If natural coordinates are used, the following expression is obtained. For two Gaussian distributions with common covariance \(P_1\), the Kullback-Leibler divergence becomes:
\begin{equation}
D_{\mathrm{KL}}(p_i\|p_j)
=
\frac12
(\eta_j-\eta_i)^\top
P_1
(\eta_j-\eta_i)
=
\frac12 d_M^2(p_i,p_j).
\end{equation}

Therefore, on the submanifold of Gaussian distributions with fixed covariance, the informational divergence is proportional to the square of the geometric distance induced by the Fisher metric.

\section{Simulation I: Common Posterior Covariance}
The framework described in the previous sections is illustrated through a simulation exercise. 
We consider a portfolio composed of \(N\) borrowers. Each borrower is associated with an individual latent variable \(\theta\), which, by construction, is not directly observable.

With reference to the prior distribution of \(\theta \mid z\), we assume that it is described by a bivariate Gaussian distribution with \mbox{borrower-specific} mean and constant variance-covariance matrix. The mean of the creditworthiness component is determined by the acceptance score, while the mean of the fragility component is described by the corresponding fragility score.

We assume that higher values of the first score correspond to higher expected creditworthiness, while higher values of the second score correspond to greater expected financial fragility.

Once the prior distribution has been simulated, we construct the behavioural monitoring scores, that is, the likelihood, associated with creditworthiness, the first component, and financial fragility, the second component.

Given the prior distribution and the likelihood, we derive the posterior distribution of the latent variables representing creditworthiness and fragility. This posterior distribution represents the bank's updated knowledge of the latent state conditional on the observed data.

Once the expectation parameters associated with the posterior distributions have been obtained, the corresponding natural coordinates are derived.
In banking practice, it is common to segment the population on the basis of observed \textit{scores}. 

For instance, suppose that the bank classifies as high-quality borrowers those with a sufficiently high behavioural risk score and a sufficiently low financial fragility score.

Throughout the simulations, the first score is oriented so that higher values correspond to better credit quality, whereas the second score is oriented so that higher values correspond to greater financial fragility and lower affordability.

Borrowers are assigned to the low-quality segment if either the risk score is below or equal to a given threshold, or the fragility score is above or equal to a given threshold.

Once borrowers have been assigned to segments of different credit quality, it is of interest to compute the Kullback-Leibler divergences between the posterior distributions, using natural coordinates.
The resulting algorithm is presented below.

\begin{algorithm} 
\caption{Baseline simulation with common posterior covariance} 
\begin{enumerate} 
\item Generate \(N\) initial borrower covariates \(z_i\). \item Define prior means \(m_i=m(z_i)\) and a common covariance matrix \(P\). 
\item Draw latent states \(\theta_i\sim N(m_i,P)\). \item Generate observed monitoring scores \(x_i=A\theta_i+\varepsilon_i\), with \(\varepsilon_i\sim N(0,\Sigma)\). 
\item Compute the posterior covariance $P_1$ and posterior means \(m_{1,i}\). 
\item Transform posterior means into natural coordinates \(\eta_i=\Lambda m_{1,i}\). 
\item Assign borrowers to qualitative segments based on observed scores. 
\item Compute KL divergences between segment representative posterior distributions.
\end{enumerate}
\end{algorithm}

The resulting distances do not measure differences between individual borrowers. Rather, they quantify the informational divergence between the representative distributions associated with the average profiles of the three segments. The following table reports the Kullback-Leibler divergence computed in the system of natural coordinates.

\begin{table}[htbp] 
\centering 
\caption{ KL divergences over \textit{high}, \textit{medium}, \textit{low}} 
\begin{tabular}{l S[table-format=2.0] S[table-format=1.0] S[table-format=2.0]} 
\toprule & {high} & {medium} & {low} \\ 
\midrule
High   & 0.00 & 7.84 & 27.52 \\ 
Medium & 7.84 & 0.00 & 6.00 \\
Low    & 27.52 & 6.00 & 0.00 \\ 
\bottomrule 
\end{tabular} 
\end{table}

It is useful to examine the distance, in natural coordinates, between the distributions associated with the average profiles of high-, medium-, and low-quality borrowers. During the monitoring process, the evolution of these distances provides the bank with information on the dynamics of its credit portfolios and on the possible migration of borrowers across risk segments.

\section{Model Extensions}

We now remove the assumption, adopted in the previous sections, that the posterior variance-covariance matrix is constant across borrowers. This assumption allowed us to study a submanifold of the Gaussian family parametrized exclusively by posterior means, while keeping the uncertainty structure fixed. In that case, the Fisher metric reduced to the Mahalanobis metric, and the Kullback-Leibler divergence between two posterior distributions with the same covariance was proportional to the square of the geometric distance.

In the general case, instead, we assume that the prior variance-covariance matrix depends on the borrower's initial information:
\begin{equation}
    \theta \mid z \sim N(m(z),P(z)).
\end{equation}

In this way, in addition to differing in the mean position of the prior distribution, borrowers also differ in the degree of uncertainty assigned by the financial intermediary to the latent variables Creditworthiness and Fragility, as well as in the dependence structure between them. The matrix \(P(z)\) is therefore not merely a technical parameter, but describes the initial confidence that the bank assigns to its informational representation of the borrower.

In the linear-Gaussian model considered in the previous sections, the posterior distribution remains Gaussian:
\begin{equation}
    \theta \mid x,z \sim N(m_1(z,x),P_1(z)),
\end{equation}
where
\begin{equation}
    P_1(z)
    =
    \left(
    P(z)^{-1}+A^\top\Sigma^{-1}A
    \right)^{-1},
\end{equation}
and
\begin{equation}
    m_1(z,x)
    =
    P_1(z)
    \left[
    P(z)^{-1}m(z)+A^\top\Sigma^{-1}x
    \right].
\end{equation}

Thus, each borrower is no longer represented only by a posterior mean, but by the pair:
\begin{equation}
    (m_1(z,x),P_1(z)).
\end{equation}

From this point of view, the relevant statistical manifold is no longer the submanifold of Gaussian distributions with fixed covariance, but rather the full manifold of bivariate Gaussian posterior distributions. Each borrower \(i\) is therefore associated with a point of the form:
\begin{equation}
    p_i=N(m_{1,i},P_{1,i}),
\end{equation}
where \(m_{1,i}\) represents the bank's updated mean knowledge and \(P_{1,i}\) represents the residual uncertainty structure associated with that knowledge.

\subsection{The Fisher metric on the posterior Gaussian manifold}

When the posterior covariance matrix varies from borrower to borrower, the geometry of the manifold must account for two different sources of variation. The first concerns the position of the distribution, namely the vector of posterior means. The second concerns the shape of the distribution, namely the posterior variance-covariance matrix.

For the multivariate Gaussian family, the Fisher metric expressed in expectation coordinates \((m_1,P_1)\) takes the form:
\begin{equation}
    ds^2
    =
    dm_1^\top P_1^{-1}dm_1
    +
    \frac{1}{2}
    \operatorname{tr}
    \left(
    P_1^{-1}dP_1\,P_1^{-1}dP_1
    \right).
\end{equation}

The first term,
\begin{equation}
    dm_1^\top P_1^{-1}dm_1,
\end{equation}
measures the variation in the mean position of the posterior distribution. In the credit-risk context, it represents the difference in the bank's average informational state with respect to the latent dimensions of Creditworthiness and Fragility.

The second term,
\begin{equation}
    \frac{1}{2}
    \operatorname{tr}
    \left(
    P_1^{-1}dP_1\,P_1^{-1}dP_1
    \right),
\end{equation}
measures instead the variation in the posterior uncertainty structure. This component is absent in the fixed-covariance case and becomes essential when the bank assigns different levels of informational reliability to different borrowers.

This decomposition has an immediate economic interpretation. Two borrowers may be geometrically distant for two distinct reasons. They may be distant because the bank assigns them different average Creditworthiness or different average Fragility. Alternatively, they may be distant because, even for the same average assessment, the bank is much more uncertain about one borrower than about the other.

The informational distance between borrowers therefore depends not only on the difference between their posterior mean assessments, but also on the different precision with which such assessments are formulated.

\subsection{Natural coordinates in the general case}

In the previous sections, the natural coordinates of the Gaussian distribution were introduced. With reference to the posterior distribution, they are defined as:
\begin{equation}
    \Lambda_1=P_1^{-1},
\end{equation}
\begin{equation}
    \eta_1=\Lambda_1 m_1.
\end{equation}

In the fixed-covariance case, the matrix \(\Lambda_1\) was common to all borrowers and the manifold could be parametrized exclusively by the vector \(\eta_1\). When, instead, \(P_1\) varies from borrower to borrower, \(\Lambda_1\) also becomes a coordinate of the manifold.

A point on the posterior Gaussian manifold is therefore described by the pair:
\begin{equation}
    (\eta_1,\Lambda_1),
\end{equation}
where
\begin{equation}
    \eta_1 \in \mathbb{R}^2,
\end{equation}
and \(\Lambda_1\) is a symmetric positive definite \(2 \times 2\) matrix.

The potential function of the Gaussian family, expressed in posterior natural coordinates, is:
\begin{equation}
    \psi(\eta_1,\Lambda_1)
    =
    \frac{1}{2}\eta_1^\top\Lambda_1^{-1}\eta_1
    -
    \frac{1}{2}\log|\Lambda_1|
    +
    \frac{d}{2}\log(2\pi).
\end{equation}

In the present case \(d=2\), since the latent variables are Creditworthiness and Fragility.

The expectation coordinates and the posterior natural coordinates are related by:
\begin{equation}
    P_1=\Lambda_1^{-1},
\end{equation}
\begin{equation}
    m_1=\Lambda_1^{-1}\eta_1.
\end{equation}

Denoting \(P_1=\Lambda_1^{-1}\), an infinitesimal variation in the posterior mean can be obtained by differentiating the relation \(m_1=P_1\eta_1\). Since
\begin{equation}
    dP_1=-P_1\,d\Lambda_1\,P_1,
\end{equation}
we have:
\begin{equation}
    dm_1
    =
    d(P_1\eta_1)
    =
    dP_1\,\eta_1+P_1\,d\eta_1.
\end{equation}

Substituting the expression for \(dP_1\), we obtain:
\begin{equation}
    dm_1
    =
    -P_1\,d\Lambda_1\,P_1\eta_1+P_1\,d\eta_1.
\end{equation}

Since \(m_1=P_1\eta_1\), it follows that:
\begin{equation}
    dm_1
    =
    P_1(d\eta_1-d\Lambda_1\,m_1).
\end{equation}

This relation shows that, when the posterior precision \(\Lambda_1\) also varies, the variation in the posterior mean depends not only on \(d\eta_1\), but also on the variation in the precision matrix.

Similarly, starting from \(P_1=\Lambda_1^{-1}\), we obtain:
\begin{equation}
    dP_1=-P_1\,d\Lambda_1\,P_1.
\end{equation}

Substituting these relations into the Fisher metric of the Gaussian family yields the line element in posterior natural coordinates:
\begin{equation}
    ds^2
    =
    (d\eta_1-d\Lambda_1\,m_1)^\top
    P_1
    (d\eta_1-d\Lambda_1\,m_1)
    +
    \frac{1}{2}
    \operatorname{tr}
    \left(
    P_1\,d\Lambda_1\,P_1\,d\Lambda_1
    \right).
\end{equation}

Since \(m_1=P_1\eta_1\), the same expression can be written entirely in natural coordinates as:
\begin{equation}
    ds^2
    =
    (d\eta_1-d\Lambda_1\,P_1\eta_1)^\top
    P_1
    (d\eta_1-d\Lambda_1\,P_1\eta_1)
    +
    \frac{1}{2}
    \operatorname{tr}
    \left(
    P_1\,d\Lambda_1\,P_1\,d\Lambda_1
    \right),
    \qquad
    P_1=\Lambda_1^{-1}.
\end{equation}

This formula represents the natural extension, in the space of posterior natural coordinates, of the metric obtained on the fixed-covariance submanifold.

Indeed, if the posterior precision matrix is constant across borrowers, then:
\begin{equation}
    d\Lambda_1=0.
\end{equation}

In this case, the metric reduces to:
\begin{equation}
    ds^2=d\eta_1^\top P_1 d\eta_1,
\end{equation}
which coincides with the expression already obtained in Section 3. The previous metric is therefore a special case of the general metric on the posterior Gaussian manifold.

In the general case, instead, the line element contains two components:
\begin{equation}
    ds^2
    =
    \underbrace{
    (d\eta_1-d\Lambda_1\,m_1)^\top
    P_1
    (d\eta_1-d\Lambda_1\,m_1)
    }_{\text{variation in informational position}}
    +
    \underbrace{
    \frac{1}{2}
    \operatorname{tr}
    \left(
    P_1\,d\Lambda_1\,P_1\,d\Lambda_1
    \right)
    }_{\text{variation in informational precision}}.
\end{equation}

The first term measures the variation in the position of the bank's belief, taking into account the fact that precision may also change. The second term measures the variation in precision itself. In credit-risk terms, the distance between two borrowers is therefore not determined only by the difference in the posterior expected value of the latent variables Creditworthiness and Fragility, but also by the difference in the confidence with which such values are estimated.

\subsection{Local metric between two borrowers}

Let \(i\) and \(j\) be two borrowers, represented by their respective posterior Gaussian distributions:
\begin{equation}
    p_i=N(m_{1,i},P_{1,i}),
\end{equation}
\begin{equation}
    p_j=N(m_{1,j},P_{1,j}).
\end{equation}

In posterior natural coordinates, these distributions are identified by:
\begin{equation}
    (\eta_{1,i},\Lambda_{1,i}),
    \qquad
    (\eta_{1,j},\Lambda_{1,j}),
\end{equation}
where
\begin{equation}
    \Lambda_{1,i}=P_{1,i}^{-1},
    \qquad
    \eta_{1,i}=\Lambda_{1,i}m_{1,i},
\end{equation}
and
\begin{equation}
    \Lambda_{1,j}=P_{1,j}^{-1},
    \qquad
    \eta_{1,j}=\Lambda_{1,j}m_{1,j}.
\end{equation}

For small displacements on the manifold, the geometric distance between the two points can be approximated by evaluating the metric at an intermediate point. Define:
\begin{equation}
    \Delta\eta_1=\eta_{1,j}-\eta_{1,i},
\end{equation}
\begin{equation}
    \Delta\Lambda_1=\Lambda_{1,j}-\Lambda_{1,i}.
\end{equation}

Let \((\bar{\eta}_1,\bar{\Lambda}_1)\) be a local reference point, for example the midpoint in natural coordinates:
\begin{equation}
    \bar{\eta}_1=\frac{\eta_{1,i}+\eta_{1,j}}{2},
\end{equation}
\begin{equation}
    \bar{\Lambda}_1=\frac{\Lambda_{1,i}+\Lambda_{1,j}}{2}.
\end{equation}

Setting
\begin{equation}
    \bar{P}_1=\bar{\Lambda}_1^{-1},
\end{equation}
and
\begin{equation}
    \bar{m}_1=\bar{P}_1\bar{\eta}_1,
\end{equation}
we obtain the following local approximation of the geometric distance:
\begin{equation}
    d_g^2(p_i,p_j)
    \approx
    (\Delta\eta_1-\Delta\Lambda_1\,\bar{m}_1)^\top
    \bar{P}_1
    (\Delta\eta_1-\Delta\Lambda_1\,\bar{m}_1)
    +
    \frac{1}{2}
    \operatorname{tr}
    \left(
    \bar{P}_1\Delta\Lambda_1
    \bar{P}_1\Delta\Lambda_1
    \right).
\end{equation}

This expression generalizes the Mahalanobis distance used on the fixed-covariance submanifold. Indeed, if the two borrowers have the same posterior precision matrix, namely:
\begin{equation}
    \Lambda_{1,i}=\Lambda_{1,j}=\Lambda_1,
\end{equation}
then:
\begin{equation}
    \Delta\Lambda_1=0.
\end{equation}

The local distance therefore reduces to:
\begin{equation}
    d_g^2(p_i,p_j)
    =
    \Delta\eta_1^\top P_1\Delta\eta_1.
\end{equation}

Since \(\eta_1=\Lambda_1 m_1\), we have:
\begin{equation}
    \Delta\eta_1=\Lambda_1(m_{1,j}-m_{1,i}).
\end{equation}

Therefore:
\begin{equation}
    d_g^2(p_i,p_j)
    =
    (m_{1,j}-m_{1,i})^\top
    \Lambda_1
    (m_{1,j}-m_{1,i}),
\end{equation}
which coincides with the Mahalanobis distance between posterior means.

The Mahalanobis distance is therefore a special case of the distance induced by the Fisher metric, obtained when the posterior uncertainty structure is common to all borrowers.

\subsection{Kullback-Leibler divergence in the general case}

When posterior covariance matrices differ across borrowers, the Kullback-Leibler divergence between two posterior Gaussian distributions no longer reduces to the quadratic term on the means alone.

Given two distributions:
\begin{equation}
    p_i=N(m_{1,i},P_{1,i}),
\end{equation}
\begin{equation}
    p_j=N(m_{1,j},P_{1,j}),
\end{equation}
the Kullback-Leibler divergence is:
\begin{equation}
    D_{KL}(p_i\|p_j)
    =
    \frac{1}{2}
    \left[
    \operatorname{tr}(P_{1,j}^{-1}P_{1,i})
    +
    (m_{1,j}-m_{1,i})^\top P_{1,j}^{-1}(m_{1,j}-m_{1,i})
    -
    d
    +
    \log\frac{|P_{1,j}|}{|P_{1,i}|}
    \right].
\end{equation}

This expression contains four terms. The first compares the posterior covariance structures of the two distributions. The second measures the distance between posterior means, weighted by the precision of the reference distribution \(p_j\). The third is a normalization constant related to the dimension of the latent space. The fourth compares the uncertainty volumes associated with the two distributions.

In posterior natural coordinates, recalling that:
\begin{equation}
    \Lambda_{1,i}=P_{1,i}^{-1},
    \qquad
    \Lambda_{1,j}=P_{1,j}^{-1},
\end{equation}
and that:
\begin{equation}
    m_{1,i}=\Lambda_{1,i}^{-1}\eta_{1,i},
    \qquad
    m_{1,j}=\Lambda_{1,j}^{-1}\eta_{1,j},
\end{equation}
the divergence can be written as:
\begin{equation}
    D_{KL}(p_i\|p_j)
    =
    \frac{1}{2}
    \left[
    \operatorname{tr}(\Lambda_{1,j}\Lambda_{1,i}^{-1})
    +
    (\Lambda_{1,j}^{-1}\eta_{1,j}-\Lambda_{1,i}^{-1}\eta_{1,i})^\top
    \Lambda_{1,j}
    (\Lambda_{1,j}^{-1}\eta_{1,j}-\Lambda_{1,i}^{-1}\eta_{1,i})
    -
    d
    +
    \log\frac{|\Lambda_{1,i}|}{|\Lambda_{1,j}|}
    \right].
\end{equation}

This formula shows that, when \(P_1\) varies from borrower to borrower, the informational divergence simultaneously compares:
\begin{enumerate}
    \item the different mean positions of the posterior distributions;
    \item the different precision assigned to the latent variables;
    \item the different uncertainty volumes associated with the borrowers.
\end{enumerate}

Unlike in the common-covariance case, the Kullback-Leibler divergence is not symmetric:
\begin{equation}
    D_{KL}(p_i\|p_j)\neq D_{KL}(p_j\|p_i).
\end{equation}

This asymmetry is consistent with the informational interpretation of the divergence. Measuring how much information is lost by using \(p_j\) in place of \(p_i\) is not equivalent to measuring how much information is lost by using \(p_i\) in place of \(p_j\).

In the credit-risk context, this means that taking one borrower as an informational reference and comparing a second borrower against it does not necessarily produce the same result as the reverse comparison. This property may be relevant in the construction of directional measures of informational migration, for example when one wants to assess whether a borrower is moving closer to the average profile of a higher-risk or lower-risk segment.

If a symmetric measure is desired, one may use the Jeffreys divergence:
\begin{equation}
    J(p_i,p_j)
    =
    D_{KL}(p_i\|p_j)
    +
    D_{KL}(p_j\|p_i).
\end{equation}

It preserves the informational content of the Kullback-Leibler divergence, while removing the directionality of the comparison.

\subsection{Credit-risk interpretation of the extension}

The extension to the case \(P=P(z)\) enriches the interpretation of credit risk as an informational state of the intermediary. In the previous sections, the bank differentiated borrowers only according to the mean position of their posterior distributions. In other words, borrowers could have different expected Creditworthiness and different expected Fragility, but the bank assigned the same degree of posterior uncertainty to all of them.

In the general case, instead, the bank can distinguish two informational dimensions:
\begin{equation}
    \text{expected level of risk}
\end{equation}
and
\begin{equation}
    \text{uncertainty associated with the risk assessment}.
\end{equation}

This distinction is relevant in banking practice. Two borrowers may have the same expected mean value of Creditworthiness and Fragility, but be very different from an informational point of view. For a borrower with a long credit history, rich information, and stable behaviour, the bank may have a concentrated posterior distribution. For a new borrower, with little available information or contradictory signals, the posterior distribution may be much more dispersed.

In the first case, the bank assigns a relatively precise assessment. In the second case, it assigns a more uncertain assessment. The geometry of the Gaussian manifold allows both situations to be represented.

From this perspective, credit monitoring does not consist only in observing whether the borrower improves or deteriorates in average terms, but also in assessing how the bank's uncertainty about the borrower changes. An observed score can in fact produce two distinct effects:
\begin{equation}
    m_{1,i} \longrightarrow m_{1,i}',
\end{equation}
namely a displacement of the posterior mean, and
\begin{equation}
    P_{1,i} \longrightarrow P_{1,i}',
\end{equation}
namely a variation in the posterior uncertainty structure.

The first effect concerns the change in the borrower's average assessment. The second concerns the change in the confidence with which that assessment is formulated. A borrower may deteriorate because expected fragility increases, but may also become more problematic because the bank becomes more uncertain about the borrower's actual financial condition.

The geometric distance between informational states therefore allows credit monitoring to be described as a trajectory on the manifold of posterior distributions. At each observation date, the borrower occupies a point on the manifold. The arrival of new scores induces a displacement of that point, modifying both the mean position and the uncertainty structure.

Formally, for a borrower observed over time \(t=0,1,2,\ldots\), one obtains a sequence of distributions:
\begin{equation}
    p_{i,t}=N(m_{1,i,t},P_{1,i,t}).
\end{equation}

The borrower's informational dynamics can then be represented as a trajectory:
\begin{equation}
    p_{i,0}
    \longrightarrow
    p_{i,1}
    \longrightarrow
    p_{i,2}
    \longrightarrow
    \cdots
\end{equation}
on the Gaussian manifold. The geometric length of this trajectory can be interpreted as the intensity of the informational change observed by the bank.

In this sense, credit risk is no longer merely a pointwise probability or membership in a rating class, but becomes an evolutionary dynamics of the financial intermediary's beliefs.

\subsection{Probability of default as a functional of the posterior distribution}

The proposed approach does not eliminate the probability of default, but reinterprets it as a functional of the subjective posterior distribution. In a traditional model, the probability of default is often treated as a quantity directly associated with the borrower. In the present framework, instead, it derives from the bank's informational state over the latent variables Creditworthiness and Fragility.

Let \(q(\theta)\) be a function that associates each latent configuration \(\theta\) with a conditional probability of default. For example, one may assume a logistic specification:
\begin{equation}
    q(\theta)
    =
    \mathbb{P}(D=1\mid \theta)
    =
    \frac{1}{1+\exp[-(\alpha+\beta^\top\theta)]}.
\end{equation}

The probability of default associated with the borrower, conditional on the observed information, is then given by:
\begin{equation}
    PD(x,z)
    =
    \mathbb{E}_{p(\theta\mid x,z)}
    \left[
    q(\theta)
    \right].
\end{equation}

Explicitly:
\begin{equation}
    PD(x,z)
    =
    \int q(\theta)p(\theta\mid x,z)d\theta.
\end{equation}

This expression highlights that the probability of default is not an ``objective'' characteristic of the borrower, but a summary of the bank's informational state. It depends on the posterior distribution \(p(\theta\mid x,z)\), and therefore both on the mean \(m_1(z,x)\) and on the variance-covariance matrix \(P_1(z)\).

In particular, two borrowers with the same posterior mean may have different probabilities of default if the function \(q(\theta)\) is nonlinear and if their respective posterior covariance matrices differ. Uncertainty is therefore not a secondary element, but may directly affect the synthetic measure of risk.

In operational terms, the probability of default can be viewed as a one-dimensional projection of the posterior distribution. Information geometry, instead, preserves the full content of the distribution, allowing borrowers and segments to be compared not only on the basis of expected risk, but also on the basis of the uncertainty structure.

\subsection{Customer segments and representative distributions}

The extension to the case \(P=P(z)\) also modifies the comparison between customer segments. In Section 4, under common covariance, the high-, medium-, and low-credit-quality segments were compared through the Kullback-Leibler divergence between representative distributions of average profiles. In that case, the divergence depended only on the distance between means, since the posterior covariance matrix was common.

When, instead, each borrower has an individual matrix \(P_{1,i}\), the representative distribution of a segment must summarize both the individual posterior means and covariance matrices. For a segment \(G\), one may define a representative Gaussian distribution:
\begin{equation}
    \bar{p}_G=N(\bar{m}_{1,G},\bar{P}_{1,G}),
\end{equation}
where:
\begin{equation}
    \bar{m}_{1,G}
    =
    \frac{1}{|G|}
    \sum_{i\in G}m_{1,i},
\end{equation}
and, as a first approximation,
\begin{equation}
    \bar{P}_{1,G}
    =
    \frac{1}{|G|}
    \sum_{i\in G}P_{1,i}.
\end{equation}

This choice makes it possible to represent each segment through an average distribution that incorporates both the expected profile of the borrowers belonging to the group and the average level of uncertainty associated with that profile.

The comparison between two segments \(G\) and \(H\) may therefore be performed through:
\begin{equation}
    D_{KL}(\bar{p}_G\|\bar{p}_H),
\end{equation}
or, if a symmetric measure is desired,
\begin{equation}
    J(\bar{p}_G,\bar{p}_H)
    =
    D_{KL}(\bar{p}_G\|\bar{p}_H)
    +
    D_{KL}(\bar{p}_H\|\bar{p}_G).
\end{equation}

In this way, the distance between segments measures not only the difference between borrowers who are, on average, more or less risky, but also the difference between segments for which the bank has different levels of informational certainty.

This distinction is particularly important in portfolio management. A segment may be considered risky not only because it has high average fragility or low average creditworthiness, but also because it is characterized by high uncertainty. Similarly, a segment may be considered stable not only because it presents good average indicators, but because the bank has precise and coherent information about the borrowers composing it.

\subsection{Summary of the extension}

The introduction of a borrower-dependent prior variance-covariance matrix allows us to move from a simplified geometry, defined on the submanifold of Gaussian distributions with fixed posterior covariance, to the full geometry of the Gaussian family.

In the first case, the distance between borrowers depends exclusively on the difference between posterior means:
\begin{equation}
    d_g^2(p_i,p_j)
    =
    (m_{1,j}-m_{1,i})^\top\Lambda_1(m_{1,j}-m_{1,i}).
\end{equation}

In the second case, the distance depends both on the difference between posterior means and on the difference between uncertainty structures:
\begin{equation}
    ds^2
    =
    dm_1^\top P_1^{-1}dm_1
    +
    \frac{1}{2}
    \operatorname{tr}
    \left(
    P_1^{-1}dP_1\,P_1^{-1}dP_1
    \right).
\end{equation}

In posterior natural coordinates, this metric becomes:
\begin{equation}
    ds^2
    =
    (d\eta_1-d\Lambda_1\,m_1)^\top
    P_1
    (d\eta_1-d\Lambda_1\,m_1)
    +
    \frac{1}{2}
    \operatorname{tr}
    \left(
    P_1\,d\Lambda_1\,P_1\,d\Lambda_1
    \right).
\end{equation}

This formulation clarifies the role of informational precision in the geometry of risk. The bank does not directly observe the borrower's latent states, but progressively updates a subjective distribution over those states. Credit risk is therefore represented by the bank's position on the manifold of posterior distributions. The scores observed during the monitoring process determine displacements on this manifold, modifying both the expected level of the latent variables and the uncertainty associated with their estimation.

The proposed extension therefore makes it possible to interpret credit monitoring as a geometric dynamics of the financial intermediary's informational state. From this perspective, risk management does not consist only in measuring a pointwise probability of default, but in analysing the evolution of the bank's beliefs about its borrowers.

\section{Simulation II: Borrower-Specific Covariance}

The previous construction assumes a common prior distribution for all individuals. In many risk applications, this assumption is restrictive, since different borrowers may exhibit different levels of uncertainty and different correlation structures between the latent variables. We therefore introduce a vector of observable covariates \[ z_i=(z_{i1},z_{i2}), \] which describes preliminary borrower-specific characteristics. These covariates determine both the prior mean \[ m_i = m(z_i), \] and the covariance matrix \[ P_i=P(z_i). \] In the bivariate case considered here, the prior mean represents the expected level of the two latent components, for example creditworthiness and fragility, while the covariance structure is parametrized as follows:

\[
P_i=
\begin{pmatrix}
e^{u_i} &
\rho_i e^{(u_i+v_i)/2}\\
\rho_i e^{(u_i+v_i)/2} &
e^{v_i}
\end{pmatrix},
\]

where \(u_i\), \(v_i\), and \(\rho_i\) depend on the covariates. The exponential parametrization ensures the positivity of the variances, 
while the hyperbolic transformation used for \(\rho_i\) ensures that the correlation remains in the interval \((-1,1)\).

Each individual is therefore associated with a Gaussian prior distribution

\[
\theta_i \sim N(m_i,P_i),
\]

and has a local geometry of the statistical space.
The Fisher metric is no longer constant, but varies across the population through the family
\(\{P_i\}_{i=1}^N\).

The observations are generated through a linear-Gaussian model

\[
x_i=A\theta_i+\varepsilon_i,
\qquad
\varepsilon_i\sim N(0,\Sigma),
\]

from which the posterior distribution is obtained:

\[
\theta_i|x_i \sim N(m_{1,i},P_{1,i}),
\]

with

\[
P_{1,i}
=
\left(
P_i^{-1}
+
A^\top \Sigma^{-1} A
\right)^{-1}
\]

and

\[
m_{1,i}
=
P_{1,i}
\left(
P_i^{-1}m_i
+
A^\top \Sigma^{-1} x_i
\right).
\]

The pair 
\(
(\eta_i,\Lambda_i)
\),
where

\[
\Lambda_i=(P_{1,i})^{-1},
\qquad
\eta_i=\Lambda_i m_{1,i},
\]

also provides the representation of the Gaussian exponential family in natural coordinates.

This generalization transforms the problem from a homogeneous statistical space into a manifold of distributions in which each borrower has an individual metric structure. The distance between population segments can therefore be studied by comparing the aggregate Gaussian distributions of the groups through Kullback-Leibler divergences and the Jeffreys distance.

Borrowers are finally grouped into qualitative classes, namely high, medium, and low quality, on the basis of the observed scores. For each group, a representative Gaussian distribution is constructed using the average of the individual posterior means and the average covariance matrix of the group. 

The Kullback-Leibler divergences between these distributions measure the informational cost associated with replacing one segment with another. The symmetrized Jeffreys divergence

\[
J(P,Q)=D_{KL}(P\|Q)+D_{KL}(Q\|P)
\]

provides a bidirectional measure of the geometric separation between segments, allowing one to quantify how distinguishable the qualitative classes are in the informational space generated by the latent variables.

\begin{algorithm}[H] 
\caption{Simulation with \mbox{borrower-specific} covariance} \label{alg:borrower_specific_covariance} 
\begin{algorithmic}[1] 
\Require Number of borrowers \(N\), matrices \(A\) and \(\Sigma\), 
functions \(m(z)\) and \(P(z)\) 
\Ensure Posterior distributions 
\(N(m_{1,i},P_{1,i})\), 
natural coordinates 
\((\eta_i^{post},\Lambda_i^{post})\), segment-level divergences
\State Generate \(N\) borrower-specific covariates 
\(z_i\) 
\State Define prior means \(m_i = m(z_i)\)
\State Construct borrower-specific covariance matrices \(P_i=P(z_i)\) 
\State Draw latent states \(\theta_i\sim N(m_i,P_i)\) 
\State Generate observed scores \(x_i=A\theta_i+\varepsilon_i\), with \(\varepsilon_i\sim N(0,\Sigma)\)
\State Compute individual posterior distributions \(N(m_{1,i},P_{1,i})\)
\State Compute natural coordinates \((\eta_i,\Lambda_i)\) 
\State Aggregate individual posterior distributions into segment-level representative Gaussian distributions 
\State Compute asymmetric Kullback-Leibler divergences and symmetric Jeffreys divergences between segments
\end{algorithmic} 
\end{algorithm}

With reference to the Kullback-Leibler divergences, the asymmetric values reported in the following table are obtained, as expected.

\begin{table}[htbp] \centering \caption{KL comparison between \textit{high}, \textit{medium}, and \textit{low} population segments} \begin{tabular}{l S[table-format=2.2] S[table-format=2.2] S[table-format=2.2]} \toprule & {High} & {Medium} & {Low} \\ \midrule High & 0.00 & 10.34 & 36.59 \\ Medium & 10.18 & 0.00 & 8.26 \\ Low & 35.81 & 8.24 & 0.00 \\ \bottomrule \end{tabular} \end{table} 

It is of interest to compare these values with those obtained from the symmetric Jeffreys divergence.

\begin{table}[htbp] \centering \caption{Jeffreys divergence comparison between \textit{high}, \textit{medium}, and \textit{low} population segments} \begin{tabular}{l S[table-format=2.2] S[table-format=2.2] S[table-format=2.2]} \toprule & {High} & {Medium} & {Low} \\ \midrule High & 0.00 & 20.52 & 72.40 \\ Medium & 20.52 & 0.00 & 16.50 \\ Low & 72.40 & 16.50 & 0.00 \\ \bottomrule \end{tabular} \end{table}

The results confirm the qualitative ordering already observed in the common-covariance simulation: the high-quality and low-quality segments are the most distant, while the medium-quality segment remains closer to both. However, the interpretation is richer in the variable-covariance setting. Segment distances now reflect not only differences in expected creditworthiness and fragility, but also differences in the uncertainty structure associated with each segment.

\section{Discussion and Empirical Outlook}

The purpose of this paper is theoretical and methodological. The proposed framework is intended to show how credit risk can be represented as an informational state of the financial intermediary and how Bayesian updating naturally induces a geometry on the space of posterior distributions. The simulation exercises reported in the previous sections are therefore illustrative. They are not meant to provide an empirical calibration of a credit-risk model, but rather to clarify the role of posterior beliefs, natural coordinates, Fisher geometry, and information divergences in a controlled setting.

In the present contribution, the parameters of the linear-Gaussian model, such as the matrix \(A\), the observation covariance matrix \(\Sigma\), and the prior functions \(m(z)\) and \(P(z)\), are specified exogenously. Similarly, the functions \(u(z)\), \(v(z)\), and \(\rho(z)\), which determine the borrower-specific prior covariance matrix, are chosen in order to illustrate how heterogeneous uncertainty can be incorporated into the geometry of credit risk.

A natural continuation of this work is the development of an empirical version of the model based on anonymized banking data. In such a setting, the initial borrower information \(z\), the observed monitoring scores \(x\), and the subsequent credit outcomes could be used to estimate the parameters of the Bayesian model. In particular, one may estimate the mapping from borrower characteristics to prior means and covariance matrices, the observation equation linking latent creditworthiness and fragility to observed scores, and the uncertainty structure associated with different borrower profiles.

From a statistical point of view, this would allow the functions
\[
m(z), \qquad P(z), \qquad A, \qquad \Sigma
\]
to be inferred from data rather than imposed in the simulation design. Moreover, if default or deterioration events are observed, the probability of default could be modelled as a functional of the posterior distribution, for example through a logistic specification of the form
\[
q(\theta)=\mathbb{P}(D=1\mid\theta).
\]
This would make it possible to connect the information-geometric representation proposed in this paper with standard credit-risk quantities such as default probabilities, rating migrations, and portfolio-level risk measures.

The empirical extension would also allow one to study the dynamics of posterior distributions over time. For each borrower \(i\), repeated observations would generate a sequence of posterior distributions
\[
p_{i,0},p_{i,1},p_{i,2},\ldots,
\]
which can be interpreted as a trajectory on the statistical manifold. The geometric length, direction, and curvature of such trajectories could provide additional indicators of informational change, deterioration, stabilization, or migration across portfolio segments.

Therefore, the present paper should be viewed as a conceptual and mathematical foundation. Its main contribution is to formulate credit risk as a geometry of posterior beliefs. A subsequent empirical paper may then address calibration, estimation, validation, and comparison with traditional score-based or probability-of-default-based approaches.

A dynamic extension of the framework could be formulated as a state-space model, in which the latent creditworthiness and fragility variables evolve over time and are sequentially updated as new monitoring scores become available.

\section{Conclusion}

This paper has proposed an information-geometric interpretation of credit risk based on Bayesian posterior beliefs. Rather than treating credit risk as a pointwise attribute of the borrower, the framework represents the financial intermediary's knowledge as a probability distribution over latent dimensions of creditworthiness and financial fragility.

Under a linear-Gaussian specification, posterior distributions belong to the Gaussian exponential family and can therefore be studied as points on a statistical manifold endowed with the Fisher information metric. In the common-covariance case, the induced geometry reduces to the Mahalanobis metric, while in the borrower-specific covariance case the geometry incorporates both differences in posterior means and differences in uncertainty structures.

The simulation exercises illustrate how Bayesian updating generates posterior distributions, natural coordinates, and information divergences between borrowers and portfolio segments. In particular, the Kullback-Leibler and Jeffreys divergences provide measures of informational separation between representative segment distributions.

The proposed approach offers a conceptual bridge between Bayesian credit-risk modelling, information geometry, and portfolio monitoring. 

Future work will focus on the empirical estimation of the model using anonymized banking data, with the aim of calibrating the latent structure, validating the resulting posterior dynamics, and assessing the operational relevance of geometric indicators for credit-risk management.

\printbibliography

\appendix

\section{Reproducibility and Ancillary Code}

All simulations in this paper are based on synthetic data generated with a fixed random seed. The numerical results reported in the tables can be reproduced by running the accompanying R scripts, which are provided as ancillary files to this submission. The scripts require the \texttt{MASS} package and can be executed independently.

\end{document}